\documentclass[fleqn,twoside]{article}
\usepackage[headings]{espcrc2}

\readRCS
$Id: espcrc2.tex,v 1.2 2004/02/24 11:22:11 spepping Exp $
\usepackage{graphicx}
\usepackage[figuresright]{rotating}

\def\bbuildrel#1_#2^#3%
{\mathrel{\mathop{\kern 0pt#1}\limits_{#2}^{#3}}}

\newcommand{\ovl}[1]{\overline{#1}}
\newcommand{\ice}[1]{\relax}
\newcommand{\re}[1]{(\ref{#1})}

\newcommand{\beq}{\begin{equation}}
\newcommand{\eeq}{\end{equation}}
\newcommand{\bea}{\begin{eqnarray}}
\newcommand{\eea}{\end{eqnarray}}

\newcommand{\ba}{\begin{array}}
\newcommand{\ea}{\end{array}}

\newcommand{\MSbar}{\overline{\mbox{MS}}}
\newcommand{\msbar}{{\scriptsize \overline{\rm MS}}}

\newcommand{\os}{{\scriptsize \rm OS}}

\newcommand{\BreakI}{ \right. \nonumber \\ &{}& \left. }

\let\*=\,

\newcommand{\vep}{\varepsilon}

\newcommand{\AmS}{{\protect\the\textfont2
  A\kern-.1667em\lower.5ex\hbox{M}\kern-.125emS}}

\title{
{
 \vspace*{-14mm}
\centerline{\normalsize\hfill SFB/CPP-08-47}
\centerline{\normalsize\hfill TTP08-27}
\centerline{\normalsize\hfill BNL-HET-08/16}
{}}
\vspace{3mm}
\vspace{1cm}
New Results in Four and Five Loop
QED  calculations
\thanks{
Talk presented at 9th DESY Workshop on Elementary Particle Theory:
Loops and Legs in Quantum Field Theory, Sondershausen, Germany, 20-25 April  2008.
}}

\author{P.~A.~Baikov\address{Institute of Nuclear Physics,
        Moscow State University, \\
        Moscow~~119992, Russia}%
       ,
        K.G.Chetyrkin\address[KUNI]{Institut f\"ur Theoretische Teilchenphysik, \\
        Universit\"at Karlsruhe, D-76128 Karlsruhe, Germany
        }%
\thanks{On leave from Institute for Nuclear Research
of the Russian Academy of Sciences, Moscow, 117312, Russia.} 
            and 
C. Sturm\address{
Physics Department,
             Brookhaven National Laboratory, \\
             Upton,
             New York 11973, U.S.A
}
}

\runtitle{2-column format camera-ready paper in \LaTeX}
\runauthor{K.G. Chetyrkin}

\begin{document}

\begin{abstract}

We report on  two recent 
multiloop results  in QED:
(i) the four-loop corrections to the conversion relations between the
QED charge renormalized in the on-shell and  $\MSbar$ schemes;
(ii) analytical evaluation of  a class of
  asymptotic contributions to the muon anomaly at five-loops.

\vspace{1pc}
\end{abstract}

\maketitle

\section{Introduction}

The study of the anomalous magnetic moment of the muon $a_\mu$ 
is a long-standing challenge for both theory and experiment.
It  has been measured with impressive accuracy at the level of 0.5 parts per
million~\cite{Bennett:2006fi}: $a^{\mbox{\tiny exp}}_\mu=11 659 2080(63)
\cdot 10^{-11}$.  From theory side the anomalous magnetic moment has
been studied in great detail through the computation of higher order
corrections (see, e.g. reviews
\cite{Melnikov:2006sr,Jegerlehner:2007xe,Miller:2007kk}).
These higher order corrections to $a^{theo}_\mu$ are
basically classified into three classes: pure QED, electroweak and
hadronic contribution.

A discussion of the electroweak and hadronic corrections can be found
in Ref.~\cite{PDG} and references therein.  Within this work we
consider higher order corrections to the pure QED part. Starting from
two-loop, diagrams with internal fermion-loops can arise, where the
fermion-type of the internal loop can in general be different from the
external muon. The one-loop~\cite{Schwinger:1951nm} and two-loop
\cite{PhysRev.77.536,Petermann:1957hs,PhysRev.107.328,PhysRev.105.1930,%
PhysRev.105.1931,Elend:1966aa,Erickson:1968aa}
contributions have been computed more than 50 years ago. The three-loop
order has been computed numerically and
analytically~\cite{Kinoshita:1995ym,%
Laporta:1991zw,Laporta:1992pa,Laporta:1993qw,Laporta:1994yi,Laporta:1995vp,Laporta:1996mq} %

The complete calculation of hundreds of diagrams contributing to the muon
anomaly at four loop order is only possible by numerical integration,
which was performed in a remarkable long-term effort by Kinoshita and his
collaborators \cite{Kinoshita:2004wi,Kinoshita:2005sm}.
Even some numerically  important  five-loop diagrams 
have been computed by now \cite{Nio:2007zz}.

Accurate numerical calculation of highly divergent  multiloop  Feynman amplitudes
is highly nontrivial task. The analytical calculation (of even a very particular 
class of  diagrams)
 should be quite  useful as an independent check {\em in both directions!}
(see e.g.  instructive examples in 
Refs.~\cite{Faustov:1990zs,Kinoshita:1990ur,Kinoshita:1990wp,Broadhurst:1992za,Kinoshita:2004wi,Baikov:1995ui,Kinoshita:2005zr}).

\begin{figure}[!ht]
\begin{center}
\begin{minipage}{1.8cm}
\includegraphics[bb=108 562 224 687,width=1.8cm]{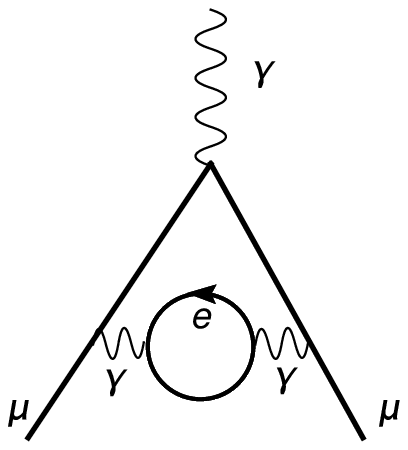}\\[-1.2cm]
\begin{center}
\vspace{0.5cm}
{\tiny{eVP}}
\end{center}
\end{minipage}
\hspace{1cm}
\begin{minipage}{2.0cm}
\includegraphics[bb=88 559 240 687,width=2.0cm]{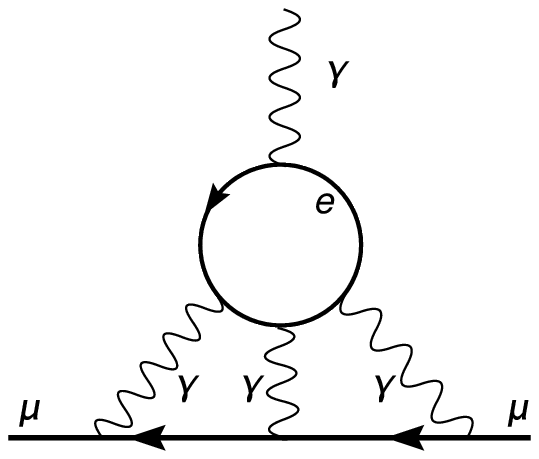}\\[-1.0cm]
\begin{center}
\vspace{0.4cm}
{\tiny{LBL}}
\end{center}
\end{minipage}
\end{center}
\vspace{-0.5cm}
\caption{\label{fig:VPLBL}Example diagrams leading to dominant
  logarithmic contributions from electron vacuum polarization (VP)
  insertions and light-by-light (LBL) scattering diagrams.}
\end{figure}

In particular  starting from two-loop there arise logarithmic
contributions of the type $\log(M_\mu/M_e)$, where $M_\mu$ is the mass
of the muon and $M_e$ the mass of the electron, respectively.  In view
of the large mass ratio $M_\mu/M_e\sim 200$ one can expect, that these
logarithms play a dominant role. The logarithmically enhanced
contributions arise from the insertion of the electron vacuum
polarization (eVP) into the first order muon vertex diagram,
 but they can also appear through
light-by-light (LBL) scattering diagrams.
 Examples for both diagram types are shown in
 Fig.~\ref{fig:VPLBL}. 

In what follows we will consider only  eVP contributions  to $a_\mu$ 
which are produced from diagrams like Fig 1 (eVP) but with the  photon propagator
receiving all possible QED perturbative corrections made of electron
loops and photon exchanges only. The general structure of eVP was
throughly studied long ago in the pioneering publications of
B. Lautrup and E. De Rafael and  R.~Barbieri and E.~Remiddi 
\cite{Lautrup:1974ic,Barbieri:1974nc}.

 
There it was  found,  that
the asymptotic part of the muon anomaly $a^{\scriptsize as}_\mu$,
which contains these logarithmic contributions originating form the
electron vacuum polarization function insertions and the mass
independent term, can be obtained with the help of the electron vacuum
polarization function in the asymptotic limit $M_e\to0$. 
The corresponding  master formula reads
\bea
\label{eq:amuasymp}
&{}&a^{\mbox{\scriptsize eVP}}_{\mu}=a^{\mbox{\scriptsize as}}_{\mu} +  \mathcal{O}\left({M_e\over    M_\mu}\right)
\nonumber
=
{\alpha \over \pi}\int_{0}^{1}\!dx\*(1-x)\*
\\
&{}&
  \left[d^{\mbox{\scriptsize
        as}}_{R}\left({x^2\over1-x}{M_\mu^2\over
        M_e^2},\alpha\right)-1\right] + \mathcal{O}\left({M_e\over    M_\mu}\right)
\label{main}
{}.
\eea
Here 
\bea
\label{eq:dR}
&{}&d^{\mbox{\scriptsize as}}_R(-q^2/M^2,\alpha)={1\over 1+\Pi^{\mbox{\scriptsize as}}(-q^2/M^2,\alpha)}
{},
\\
&{}&
\Pi(-q^2/M^2,\alpha)\bbuildrel{=\!=\!=}_{-q^2 \to \infty}^{}
\nonumber
\\
&{}&
\Pi^{\mbox{\scriptsize as}}(-q^2/M^2,\alpha) 
+ 
\mathcal{O}\left({M^2 \over  -q^2}\right)
{}.
\eea
In the above formulas 
$\Pi$ is the photon polarization operator (in QED incorporating  {\em  exactly one} fermion field),
 $M$ is the on-shell fermion mass,
and the fine structure constant  $\alpha$ is  defined 
in the classical OS-scheme  by the condition:
\begin{equation} 
\label{eq:OScharge}
\Pi(-q^2=0,M^2,\alpha) \equiv 0
{}.
\end{equation} 
Once  $\Pi^{\mbox{\scriptsize as}}$ is computed to some order in $\alpha$ the master formula 
\re{eq:amuasymp} immediately delivers the corresponding eVP contribution to $a_\mu$.

At present this technique has been applied in order to find the  complete eVP contributions to  $a_\mu$ up 
to three-loop and four-loop order \cite{Lautrup:1974ic,Barbieri:1974nc,Kinoshita:1990ur,Broadhurst:1992za}.

In our talk we extend these results by one order. As a by-product we will also
derive  the four-loop corrections to the conversion relations between the
QED charge renormalized in the on-shell and  $\MSbar$ schemes.

\section{Photon polarization operator in $\msbar$-scheme}

Thus, the main problem is to compute  the asymptotic photon polarization operator
($Q^2 \equiv -q^2$)
\beq
\label{eq:expPiOS}
\Pi^{\mbox{\scriptsize as}}(Q/M,\alpha) = 
\sum_{i} {\Pi}^{\mbox{\scriptsize as},(i)}(Q/M)\left(\frac{\alpha}{\pi}\right)^i
\nonumber
{},
\eeq
with  every 
${\Pi}^{\mbox{\scriptsize as},(i)}(Q/M)$ being, in fact\footnote{This is  a 
consequence of the Weinberg theorem \cite{Weinberg:1959nj}.},
a polynomial of order not higher than $i$ in 
$\ell_{QM} = \log(\frac{Q^2}{M^2})$.

It is very useful to consider first  the  
photon polarization operator in $\msbar$-scheme written as 
$
\ovl{\Pi}(-q^{2},{m}^2,\mu, \ovl{\alpha})
$
where $\ovl{\alpha} \equiv \alpha^{\msbar}(\mu)$ and ${m} \equiv
{m}^{\msbar}(\mu)$ are the running coupling constant and the fermion mass
in the $\MSbar$-scheme, while $\mu$ stands for the $\msbar$ renormalization point.
Indeed, the $\msbar$ polarization function $\ovl{\Pi}$ has a  smooth massless limit:
\begin{equation}
\label{eq:expm0}
\ovl{\Pi}(Q^2,m^2 =0,\ovl{\alpha}) = 
\sum_{i} \ovl{\Pi}^{(i)}(\mu^2/Q^2)\left(\frac{\ovl{ \alpha}(\mu)}{\pi} \right)^i
{}.
\end{equation}
A use of  the fundamental concept of the {\em invariant charge }  \cite{Bogolyubov:1956gh,Shirkov:1998ak} 
directly leads to  
the connection between $\ovl{\Pi}$ and  $\Pi^{\os}$ 
\begin{equation}
\label{eq:InvCharge}
{\alpha^{\os} \over 1+\Pi^{\os}(Q^2,M^2,\alpha^{\os})}=
{\ovl{\alpha} \over 1+\ovl{\Pi}(Q^2,{m}^2,\ovl{\alpha})},
\end{equation}
Eq. (\ref{eq:InvCharge})
allows one  (see e.g. \cite{Lautrup:1974ic}) to relate $\alpha^{\msbar}$ and $\alpha^{\os}$ through the $\msbar$ polarization
operator at zero momentum transfer 
\begin{equation}
\label{eq:expQ0}
\ovl{\Pi}(Q^2 =0 ,m^2,\ovl{\alpha}) = 
\sum_{i} \ovl{\Pi}^{(i)}(\mu^2/m^2)\left(\frac{\ovl{ \alpha}(\mu)}{\pi} \right)^i.
\end{equation}

Thus, to compute the $\Pi^{as}$ at four loops we need to know the
$\msbar$ polarization function $\ovl{\Pi}$ in the massless and
momentum-less limits as well as the relation between on-shell and
$\msbar$ masses of a fermion in QED at three loops. The latter,
fortunately, is available from \cite{Chetyrkin:1999qi,Melnikov:2000qh}.

\subsection{Massless limit of $\ovl{\Pi}$ }

We have computed $\ovl{\Pi}^{(4)}(\mu^2/Q^2)$ using   the parallel version of {\tt{FORM}}
\cite{Vermaseren:2000nd,Fliegner:1999jq,Fliegner:2000uy}.
The contributing diagrams were first generated  
with the package {\tt{QGRAF}} \cite{Nogueira:1991ex}.
The  reduction to master integrals was performed 
with the help of an auxiliary integral representation
\cite{Baikov:tadpoles:96}. The result reads ($\ell_{\mu Q} = \log(\frac{\mu^2}{Q^2})$):
\begin{eqnarray}
\lefteqn{\ovl{\Pi}^{(4)}(\mu^2/Q^2)= 
\frac{1075825}{373248}
-\frac{13}{8640} \,\pi^4
+\frac{13051}{2592}  \,\zeta_{3}
 }
\nonumber\\
&-&\frac{5}{3}  \,\zeta_3^2
+\frac{45}{32}  \,\zeta_{5}
-\frac{35}{4}  \,\zeta_{7}
\nonumber\\
&{+}&l_{\mu Q}
\left[
\frac{9403}{10368}
+\frac{23}{108}  \,\zeta_{3}
-\frac{5}{3}  \,\zeta_{5}
\right]
\nonumber\\
&{+}&l_{\mu Q}^2
\left[
\frac{19}{144}
-\frac{1}{9}  \,\zeta_{3}
\right]
+l_{\mu Q}^3
 \frac{1}{108}
{}.
\label{PiMS4}
\end{eqnarray}
\subsection{Momentum-less limit of  $\ovl{\Pi}$}
The limit $q^2\to0$ leads to the evaluation of massive tadpole
diagrams.  Their computation has been performed with
{\tt{FORM}}~\cite{Vermaseren:2000nd,Vermaseren:2002rp,Tentyukov:2006ys}
based programs. All appearing tadpole diagrams have been reduced to
master integrals with the help of Laporta's
algorithm~\cite{Laporta:1996mq,Laporta:2001dd}. The arising
polynomials in the space-time dimension $d=4-2\*\vep$ have been
simplified with the program {\tt{FERMAT}}~\cite{Lewis}. The remaining
master integrals are known analytically to sufficient high order in
$\vep$ and have been taken from
Refs.~\cite{Laporta:2002pg,Chetyrkin:2004fq,Kniehl:2005yc,Schroder:2005va,Schroder:2005db,Chetyrkin:2006dh,Bejdakic:2006vg,Kniehl:2006bf,Kniehl:2006bg}.
The following result for $\ovl{\Pi}^{(4)}$ at $Q=0$
was found 
\begin{eqnarray}
\lefteqn{\ovl{\Pi}^{(4)} (Q=0) =
-\frac{24254383}{9331200}
+\frac{69437}{86400} \,\pi^4
\label{tadpole}
 }
\\
\nonumber
&-&\frac{1780741}{43200}  \,\zeta_{3}
+\frac{10087}{480}  \,\zeta_{5}
-\frac{106}{675} \,\pi^4\mathrm{ln}\, 2\,
\\
\nonumber
&+&\frac{2227}{720} \,\pi^2\mathrm{ln}^2\,2\,
-\frac{32}{135} \,\pi^2\mathrm{ln}^3\,2\,
-\frac{2227}{720} \mathrm{ln}^4\,2\,
\\
\nonumber
&+&\frac{32}{225} \mathrm{ln}^5\,2\,
-\frac{2227}{30}  \,a_4
-\frac{256}{15}  \,a_5
\nonumber\\
\nonumber
&{+}&
\left[
-\frac{9383}{10368}
-\frac{29}{48}  \,\zeta_{3}
\right]\,\ell_{\mu m}
-\frac{25}{216} \ell_{\mu m}^2
-\frac{1}{108}\ell_{\mu m}^3
{}.
\label{Pi0MS4}
\end{eqnarray}
Here 
$\ell_{\mu m}=\log(\frac{\mu^2}{m^2})$, 
$\zeta_n = \zeta(n)$ is Riemann's zeta function and
$a_n = {\rm Li}_n(1/2) = \sum_{i=1}^{\infty} 1/(2^i i ^n)$.
%
\section{Conversion formulas for $\alpha$}
Let us define the conversion factor $C_{\ovl{\alpha}\alpha}$, which
converts the fine structure constant $\ovl{\alpha}$ in the  $\MSbar$-scheme
into $\alpha$ in OS-scheme: $\ovl{\alpha} = C_{\ovl{\alpha}\alpha}\,{\alpha}$:
\begin{equation} 
C_{\ovl{\alpha}\alpha} = 1+ \sum_{i \ge 1}C_{\ovl{\alpha}\alpha}^{(i)} 
\left(\frac{\alpha}{\pi} \right)^i.
\label{eq:expCabara}
\end{equation}
A use of eq. (\ref{tadpole}) directly leads to  our  result for $C_{\alpha\ovl{\alpha}}^{(4)}$ 
(for brevity we  skip the  lower order  expressions for $C_{\alpha\ovl{\alpha}}^{(i)}$ with $i=1,2,3$
which can be found in \cite{Broadhurst:1991fi})

\begin{eqnarray}
\lefteqn{C_{\ovl{\alpha}\alpha}^{(4)}(\mu^2/M^2)=  
\frac{14327767}{9331200} 
+\frac{8791}{3240} \,\pi^2
+\frac{204631}{259200} \,\pi^4
\nonumber
}
\\&{}&
-\frac{175949}{4800}  \,\zeta_{3}
+\frac{1}{24} \,\pi^2 \,\zeta_{3}
+\frac{9887}{480}  \,\zeta_{5}
\nonumber
\\&{}&
-\frac{595}{108} \,\pi^2\mathrm{ln}\, 2\, 
-\frac{106}{675} \,\pi^4\mathrm{ln}\, 2\, 
+\frac{6121}{2160} \,\pi^2\mathrm{ln}^2\,2\,
\nonumber
\\&{}&
-\frac{32}{135} \,\pi^2\mathrm{ln}^3\,2\,
-\frac{6121}{2160} \mathrm{ln}^4\,2\,
+\frac{32}{225} \mathrm{ln}^5\,2\,
\nonumber
\\&{}&
\phantom{+ }
-\frac{6121}{90}  \,a_4
-\frac{256}{15}  \,a_5
\nonumber
\nonumber\\
&{+}&\,\ell_{\mu M}\,
\left[
-\frac{383}{31104} 
+\frac{23}{108} \,\pi^2
-\frac{41}{144}  \,\zeta_{3}
\BreakI
\phantom{+\,\ell_{\mu M}\,}
-\frac{2}{9} \,\pi^2\mathrm{ln}\, 2\, 
\right]
\ice{
&{+}&\,\ell_{\mu M}\,
\left[
-\frac{383}{31104} 
+\frac{23}{108} \,\pi^2
-\frac{41}{144}  \,\zeta_{3}
-\frac{2}{9} \,\pi^2\mathrm{ln}\, 2\, 
\right]
}
\nonumber\\
&{+}&
 \frac{43}{144}\,\ell_{\mu M}^2
{+}
 \frac{13}{108}\,\ell_{\mu M}^3
{+}
 \frac{1}{81}\,\ell_{\mu M}^4
{}.
\label{bsMSfromOS4}
\end{eqnarray}
Note that in the process of deriving (\ref{bsMSfromOS4}) one needs
also $\ovl{\Pi}^{(i)}(\mu^2/m^2)$ and $\ovl{\Pi}^{(i)}(\mu^2/Q^2)$ for
$i=1,2,3$. We have not written the corresponding results explicitly to
save space; the reader could find them e.g. in
Ref.~\cite{Broadhurst:1992za}.

\section{Asymptotic photon polarization operator in on-shell scheme}
The  combined use of eqs. (\ref{PiMS4},\ref{Pi0MS4},\ref{bsMSfromOS4}) 
immediately leads us to the 
the following expression for the four-loop contribution to the asymptotic
photon polarization\footnote{
All terms proportional to $\ell$  in  (\ref{PiOs4}) could be derived 
in  easy way from  renormalization group arguments as explained in \cite{Lautrup:1974ic}. 
In addition,  the process of 
constructing $d_R^{\mbox{as}}$ one should invert the power series for
$(1+ \Pi^{\mbox{as}})$, which also produces {\em remarkably
easy-to-compute} factorizable fourth order contributions to $d_R^{\mbox{as}}$ like
$\alpha_s^4 \, \Pi^{{as},(2)}\,\Pi^{{as},(2)}$ and so  on 
(see e.g.  \cite{Kataev:2006gx} and references therein).
However,  to include power suppressed terms of order $(M/Q)^n$ to  $\Pi^{\mathrm{OS}}$
is much less trivial even for   the factorizable  contributions \cite{Czarnecki:1998rc,Aguilar:2008qj}.
}
\begin{eqnarray}
\lefteqn{\Pi^{\mbox{\scriptsize as},(4)}(Q/M) =   
\frac{5132143}{1555200} 
-\frac{8791}{3240} \,\pi^2
-\frac{205021}{259200} \,\pi^4
}
\nonumber\\
&+&\frac{5281673}{129600}  \,\zeta_{3}
-\frac{1}{24} \,\pi^2 \,\zeta_{3}
-\frac{5}{3}  \,\zeta_3^2
-\frac{2303}{120}  \,\zeta_{5}
\nonumber\\
&{}&
-\frac{35}{4}  \,\zeta_{7}
+\frac{595}{108} \,\pi^2\mathrm{ln}\, 2\, 
+\frac{106}{675} \,\pi^4\mathrm{ln}\, 2\, 
\nonumber\\
&-&\frac{6121}{2160} \,\pi^2\mathrm{ln}^2\,2\,
+\frac{32}{135} \,\pi^2\mathrm{ln}^3\,2\,
+\frac{6121}{2160} \mathrm{ln}^4\,2\,
\nonumber\\
&{}&
-\frac{32}{225} \mathrm{ln}^5\,2\,
+\frac{6121}{90}  \,a_4
+\frac{256}{15}  \,a_5
\nonumber\\
&{+}&\,\ell_{Q M}\,
\left[
-\frac{11833}{10368} 
-\frac{23}{108}  \,\zeta_{3}
+\frac{5}{3}  \,\zeta_{5}
\right]
\nonumber\\
&{+}&\,\ell_{Q M}^2
\left[
\frac{19}{144} 
-\frac{1}{9}  \,\zeta_{3}
\right]
-\frac{1}{108} \,\ell_{Q M}^3
{}.
\label{PiOs4}
\end{eqnarray}
Note that in the process of deriving (\ref{PiOs4}) one needs
also $\ovl{\Pi}^{(i)}(\mu^2/m^2)$ and $\ovl{\Pi}^{(i)}(\mu^2/Q^2)$ for
$i=1,2,3$. We have not write the corresponding results explicitly to
save space; the reader could find them 
e.g. in  Ref.~\cite{Broadhurst:1992za}.

\section{eVP contribution to the muon anomaly at fifth order}

Combining eqs. (\ref{eq:amuasymp},\ref{PiOs4}) we arrive to our main result: the
complete eVP contribution to the muon anomaly at order $\alpha^5$
(below we write down only the new five-loop result; we have confirmed
all previously known eVP contributions to $a_\mu$ at three and four
loops)
\beq 
a_{\mu}^{\mbox{\scriptsize as}}
=\sum_{i \ge 2}a_{\mu}^{\mbox{\scriptsize as},(i)} \left(
\frac{\alpha}{\pi}\right)^i
,
\eeq
\begin{eqnarray}
\lefteqn{a_{\mu}^{\mbox{\scriptsize as},(5)} =   
-\frac{296496193}{41990400} 
+\frac{45709}{58320} \,\pi^2
+\frac{212701}{518400} \,\pi^4
}
\nonumber\\
&-&\frac{4488523}{259200}  \,\zeta_{3}
+\frac{35}{144} \,\pi^2 \,\zeta_{3}
+\frac{4}{3}  \,\zeta_3^2
+\frac{10909}{720}  \,\zeta_{5}
\nonumber
\\
&+&\frac{35}{8}  \,\zeta_{7}
-\frac{55}{24} \,\pi^2\mathrm{ln}\, 2\, 
-\frac{53}{675} \,\pi^4\mathrm{ln}\, 2\, 
\nonumber
\\
&+&\frac{6121}{4320} \,\pi^2\mathrm{ln}^2\,2\,
-\frac{16}{135} \,\pi^2\mathrm{ln}^3\,2\,
-\frac{6121}{4320} \mathrm{ln}^4\,2\,
\nonumber
\\
&+&\frac{16}{225} \mathrm{ln}^5\,2\,
-\frac{6121}{180}  \,a_4
-\frac{128}{15}  \,a_5
\nonumber\\
&{+}&\,\ell_{\mu e}\,
\left[
\frac{1416095}{279936} 
+\frac{41}{972} \,\pi^2
-\frac{1855}{432}  \,\zeta_{3}
\nonumber
\BreakI
\nonumber
\phantom{MM}
-\frac{10}{3}  \,\zeta_{5}
-\frac{2}{9} \,\pi^2\mathrm{ln}\, 2\, 
\right]
\nonumber\\
&{+}&\,\ell_{\mu e}^2
\left[
-\frac{1507}{1944} 
+\frac{8}{81} \,\pi^2
+\frac{4}{3}  \,\zeta_{3}
\right]
\nonumber\\
&-&\frac{83}{243}\,\ell_{\mu e}^3
+
 \frac{8}{81}\,\ell_{\mu e}^4
{},
\label{a_mu_5_as}
\end{eqnarray}
where 
$\ell_{\mu e } = \log(M_{\mu}/M_e)$.
Finally, numerically,   the result reads:
\beq
a_{\mu}^{\mbox{\scriptsize as},(5)}\, = \,62.2667 = 63.481_{\mathrm{NS}} - 1.21429_{\mathrm{SI}}
{}.
\label{fin}
\eeq
Eq.~(\ref{fin}) also displays the decomposition of the full  result
into two pieces, corresponding to  {\em non-singlet} and {\em singlet}
contributions to the photon polarization operator (see, e.g. \cite{ChKK:Report:1996}).

\section{Conclusion and Acknowledgment}

We hope that our main result (\ref{fin}) for the complete
fifth order eVP contribution to the muon anomaly
could be of some use for testing the complicated numerical simulations like it happened a  few times 
in the past. 


We are grateful to J.H. K\"uhn for his interest in the work 
and careful reading the manuscript. 

This 
work was supported by
the Deutsche Forschungsgemeinschaft in the
Sonderforschungsbereich/Transregio
SFB/TR-9 ``Computational Particle Physics'',  by INTAS (grant
03-51-4007) and by RFBR \mbox{(grant 08-02-01451)}.
This work was also partially supported by U.S. Department
of Energy under contract No.DE-AC02-98CH10886.
The computer  calculations were partially  performed on  the  HP XC4000  super
computer of  the  federal state Baden-W\"urttemberg at the High Performance Computing Center Stuttgart
(HLRS) under the grant ``ParFORM''.

{\em \bf Note added}. Very recently, two months after the conference,   
the result for the five-loop  eVP contribution to the
muon anomaly  from singlet diagrams have been published 
by T. Aoyama, M. Hayakawa, T. Kinoshita, M. Nio and N. Watanabe 
\cite{Aoyama:2008gy}. Their result  reads:
\bea
 &{}&A_2^{(10)}(m_\mu/m_e)(\mbox{Set} \ I(j)_{(e,e)}:\mbox{combined}) =
\nonumber
\\
&{}&
 -1.24726(12)
{}.
\label{kinoshita}
\eea
One observes a good agreement for the singlet case between the
numerical  result (\ref{kinoshita}) and our prediction (\ref{fin}). The
remaining 2.6 \% difference presumably comes from power suppressed
corrections to the asymptotic result.


\begin{thebibliography}{10}

\bibitem{Bennett:2006fi}
Muon G-2, G.W. Bennett et~al.,
\newblock Phys. Rev. D73 (2006) 072003, hep-ex/0602035.

\bibitem{Melnikov:2006sr}
K. Melnikov and A. Vainshtein,
\newblock {\bf Theory of the muon anomalous magnetic moment}, Berlin, Germany:
  Springer (2006) 176 p.

\bibitem{Jegerlehner:2007xe}
F. Jegerlehner,
\newblock Acta Phys. Polon. B38 (2007) 3021, hep-ph/0703125.

\bibitem{Miller:2007kk}
J.P. Miller, E. de~Rafael and B.L. Roberts,
\newblock Rept. Prog. Phys. 70 (2007) 795, hep-ph/0703049.

\bibitem{PDG}
W.M. Yao,
\newblock J. Phys. G 33 (2007) 1, and 2007 partial update for the 2008 edition.

\bibitem{Schwinger:1951nm}
J.S. Schwinger,
\newblock Phys. Rev. 82 (1951) 664.

\bibitem{PhysRev.77.536}
R. Karplus and N.M. Kroll,
\newblock Phys. Rev. 77 (1950) 536.

\bibitem{Petermann:1957hs}
A. Petermann,
\newblock Helv. Phys. Acta 30 (1957) 407.

\bibitem{PhysRev.107.328}
C.M. Sommerfield,
\newblock Phys. Rev. 107 (1957) 328.

\bibitem{PhysRev.105.1930}
H. Suura and E.H. Wichmann,
\newblock Phys. Rev. 105 (1957) 1930.

\bibitem{PhysRev.105.1931}
A. Petermann,
\newblock Phys. Rev. 105 (1957) 1931.

\bibitem{Elend:1966aa}
H.H. Elend,
\newblock Phys. Lett. 20 (1966) 682.

\bibitem{Erickson:1968aa}
W.G. Erickson and H.T. Liu,
\newblock UCD-CNL-81 report \mbox{} (1968) \mbox{}.

\bibitem{Kinoshita:1995ym}
T. Kinoshita,
\newblock Phys. Rev. Lett. 75 (1995) 4728.

\bibitem{Laporta:1991zw}
S. Laporta and E. Remiddi,
\newblock Phys. Lett. B265 (1991) 182.

\bibitem{Laporta:1992pa}
S. Laporta and E. Remiddi,
\newblock Phys. Lett. B301 (1993) 440.

\bibitem{Laporta:1993qw}
S. Laporta,
\newblock Phys. Rev. D47 (1993) 4793.

\bibitem{Laporta:1994yi}
S. Laporta,
\newblock Phys. Lett. B343 (1995) 421, hep-ph/9410248.

\bibitem{Laporta:1995vp}
S. Laporta and E. Remiddi,
\newblock Phys. Lett. B356 (1995) 390.

\bibitem{Laporta:1996mq}
S. Laporta and E. Remiddi,
\newblock Phys. Lett. B379 (1996) 283, hep-ph/9602417.

\bibitem{Kinoshita:2004wi}
T. Kinoshita and M. Nio,
\newblock Phys. Rev. D70 (2004) 113001, hep-ph/0402206.

\bibitem{Kinoshita:2005sm}
T. Kinoshita and M. Nio,
\newblock Phys. Rev. D73 (2006) 053007, hep-ph/0512330.

\bibitem{Nio:2007zz}
M. Nio et~al.,
\newblock Nucl. Phys. Proc. Suppl. 169 (2007) 238.

\bibitem{Faustov:1990zs}
R.N. Faustov et~al.,
\newblock Phys. Lett. B254 (1991) 241.

\bibitem{Kinoshita:1990ur}
T. Kinoshita, H. Kawai and Y. Okamoto,
\newblock Phys. Lett. B254 (1991) 235.

\bibitem{Kinoshita:1990wp}
T. Kinoshita, B. Nizic and Y. Okamoto,
\newblock Phys. Rev. D41 (1990) 593.

\bibitem{Broadhurst:1992za}
D.J. Broadhurst, A.L. Kataev and O.V. Tarasov,
\newblock Phys. Lett. B298 (1993) 445, hep-ph/9210255.

\bibitem{Baikov:1995ui}
P.A. Baikov and D.J. Broadhurst,
\newblock (1995), hep-ph/9504398.

\bibitem{Kinoshita:2005zr}
T. Kinoshita and M. Nio,
\newblock Phys. Rev. D73 (2006) 013003, hep-ph/0507249.

\bibitem{Lautrup:1974ic}
B. Lautrup and E. De~Rafael,
\newblock Nucl. Phys. B70 (1974) 317.

\bibitem{Barbieri:1974nc}
R. Barbieri and E. Remiddi,
\newblock Nucl. Phys. B90 (1975) 233.

\bibitem{Weinberg:1959nj}
S. Weinberg,
\newblock Phys. Rev. 118 (1960) 838.

\bibitem{Bogolyubov:1956gh}
N.N. Bogolyubov and D.V. Shirkov,
\newblock Nuovo Cim. 3 (1956) 845.

\bibitem{Shirkov:1998ak}
D.V. Shirkov,
\newblock (1998), hep-th/9903073.

\bibitem{Chetyrkin:1999qi}
K.G. Chetyrkin and M. Steinhauser,
\newblock Nucl. Phys. B573 (2000) 617, hep-ph/9911434.

\bibitem{Melnikov:2000qh}
K. Melnikov and T.v. Ritbergen,
\newblock Phys. Lett. B482 (2000) 99, hep-ph/9912391.

\bibitem{Vermaseren:2000nd}
J.A.M. Vermaseren,
\newblock New features of FORM{,} math-ph/0010025 \mbox{} (2000) {},
  math-ph/0010025.

\bibitem{Fliegner:1999jq}
D. Fliegner, A. Retey and J.A.M. Vermaseren,
\newblock (1999), hep-ph/9906426.

\bibitem{Fliegner:2000uy}
D. Fliegner, A. Retey and J.A.M. Vermaseren,
\newblock (2000), hep-ph/0007221.

\bibitem{Nogueira:1991ex}
P. Nogueira,
\newblock J. Comput. Phys. 105 (1993) 279.

\bibitem{Baikov:tadpoles:96}
P.A. Baikov,
\newblock Phys. Lett. B385 (1996) 404, hep-ph/9603267.

\bibitem{Vermaseren:2002rp}
J.A.M. Vermaseren,
\newblock Nucl. Phys. Proc. Suppl. 116 (2003) 343, hep-ph/0211297.

\bibitem{Tentyukov:2006ys}
M. Tentyukov and J.A.M. Vermaseren,
\newblock Comput. Phys. Commun. 176 (2007) 385, cs/0604052.

\bibitem{Laporta:2001dd}
S. Laporta,
\newblock Int. J. Mod. Phys. A15 (2000) 5087, hep-ph/0102033.

\bibitem{Lewis}
R.H. Lewis,
\newblock Fermat's User Guide, http://www.bway.net/\symbol{126}lewis/ .

\bibitem{Laporta:2002pg}
S. Laporta,
\newblock Phys. Lett. B549 (2002) 115, hep-ph/0210336.

\bibitem{Chetyrkin:2004fq}
K.G. Chetyrkin et~al.,
\newblock Eur. Phys. J. C40 (2005) 361, hep-ph/0412055.

\bibitem{Kniehl:2005yc}
B.A. Kniehl and A.V. Kotikov,
\newblock Phys. Lett. B638 (2006) 531, hep-ph/0508238.

\bibitem{Schroder:2005va}
Y. Schr{\"o}der and A. Vuorinen,
\newblock JHEP 06 (2005) 051, hep-ph/0503209.

\bibitem{Schroder:2005db}
Y. Schr{\"o}der and M. Steinhauser,
\newblock Phys. Lett. B622 (2005) 124, hep-ph/0504055.

\bibitem{Chetyrkin:2006dh}
K.G. Chetyrkin et~al.,
\newblock Nucl. Phys. B742 (2006) 208, hep-ph/0601165.

\bibitem{Bejdakic:2006vg}
E. Bejdakic and Y. Schr{\"o}der,
\newblock Nucl. Phys. Proc. Suppl. 160 (2006) 155, hep-ph/0607006.

\bibitem{Kniehl:2006bf}
B.A. Kniehl and A.V. Kotikov,
\newblock Phys. Lett. B642 (2006) 68, hep-ph/0607201.

\bibitem{Kniehl:2006bg}
B.A. Kniehl et~al.,
\newblock Phys. Rev. Lett. 97 (2006) 042001, hep-ph/0607202.

\bibitem{Broadhurst:1991fi}
D.J. Broadhurst,
\newblock Z. Phys. C54 (1992) 599.

\bibitem{Kataev:2006gx}
A.L. Kataev,
\newblock {hep-ph/0602098}  (2006), hep-ph/0602098.

\bibitem{Czarnecki:1998rc}
A. Czarnecki and M. Skrzypek,
\newblock Phys. Lett. B449 (1999) 354, hep-ph/9812394.

\bibitem{Aguilar:2008qj}
J.P. Aguilar, D. Greynat and E. De~Rafael,
\newblock Phys. Rev. D77 (2008) 093010, 0802.2618.

\bibitem{ChKK:Report:1996}
K.G. Chetyrkin, J.H. K{\"u}hn and A. Kwiatkowski,
\newblock Phys. Rep. 277 (1996) 189.

\bibitem{Aoyama:2008gy}
T. Aoyama et~al.,
\newblock (2008), hep-ph/0806.3390.

\end{thebibliography}
\end{document}